\documentclass[11pt]{article}

\usepackage{amsmath}
\usepackage{amssymb}
\usepackage{tikz}
\usepackage{pgfplots}
\usepackage{graphicx}
\usepackage{caption}
\usepackage{subcaption}
\usepackage{xcolor}
\usepackage{balance}
\allowdisplaybreaks  
\usepackage{cite}
\usepackage{algorithmicx}
\usepackage{algpseudocode}
\usepackage{algorithm}
\usepackage{subcaption}
\usepackage{hyperref}
\usepackage{slashbox}
\usepackage{thmtools}

\renewcommand{\thmcontinues}[1]{cont.}

\usepackage{array}
\usepackage{authblk}
\usepackage{setspace}
\usepackage{fullpage,etoolbox}

\newtheorem{example}{Example}

\newtheorem{theorem}{Theorem}
\newtheorem{lemma}{Lemma}

\newtheorem{proposition}{Proposition}
\newtheorem{corollary}{Corollary} 
\newtheorem{remark}{Remark}
\newtheorem{definition}{Definition} 

\newcommand{\vertiii}[1]{{\vert\kern-0.25ex\vert\kern-0.25ex\vert #1\vert\kern-0.25ex\vert\kern-0.25ex\vert}}

\newtoggle{draft}
\togglefalse{draft}

\title{Temporal Robustness  of Stochastic Signals\thanks{This research was  supported by  AFOSR  grant  FA9550-19-1-0265  and  NSF  award  CPS-2038873.} }
\author{Lars Lindemann\thanks{Lars Lindemann and Al\"ena Rodionova contributed equally.} }
\author{Al\"ena Rodionova$^\dagger$}
\author{George J. Pappas}
\affil{Department of Electrical and Systems Engineering, University of Pennsylvania}

\begin{document}

\maketitle

\begin{abstract}
We study the temporal robustness of stochastic signals. This topic  is of particular interest in interleaving processes such as multi-agent systems where communication and  individual agents  induce timing uncertainty. For a deterministic signal and a given specification, we first introduce the synchronous and the asynchronous temporal robustness to quantify the signal's robustness with respect to synchronous and asynchronous time shifts in its sub-signals. We then define the temporal robustness risk by investigating the temporal robustness of the realizations of a stochastic signal. This definition can be interpreted as the risk associated with a stochastic signal to not satisfy a specification robustly in time. In this definition, general forms of  specifications such as signal temporal logic specifications are permitted. We  show how the temporal robustness risk is estimated from data for the value-at-risk. The usefulness of the temporal robustness risk is underlined by both theoretical and empirical evidence. In particular, we provide various numerical case studies including  a T-intersection scenario in autonomous driving. 
\end{abstract}

\section{Introduction}
\label{sec:introduction}

In this paper, we are interested in analyzing the robustness of time-critical systems, i.e., systems that need to satisfy stringent real-time constraints. Examples of time-critical systems include, but are not limited to, medical devices, autonomous driving, and air traffic control.   When real-time specifications are interpreted over deterministic or stochastic signals, it is natural to study  robustness  with respect to timing uncertainty  in the corresponding signals.  

\begin{figure}
	\centering
	\includegraphics[scale=0.8]{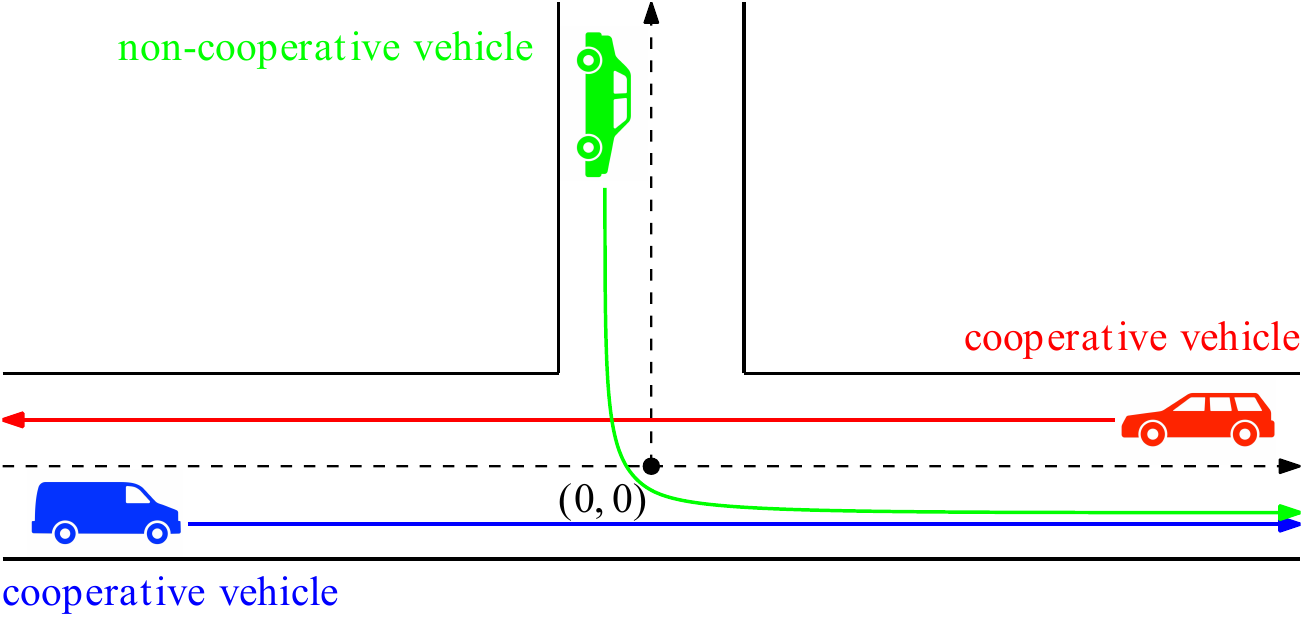}
	\caption{A T-intersection involving three cars that should pass the intersection at different times.}
	\label{ex:1_figure}
\end{figure}

For instance, consider a cooperative driving scenario in which a group of autonomous vehicles communicate to pass through a blind intersection (see Figure \ref{ex:1_figure} for an illustration of one of our case studies). Based on the transmitted information, a temporal order is assigned to the vehicles in which they must pass through the intersection. There may, however, be various reasons why an individual vehicle cannot comply with this predefined temporal order, e.g., due to delays or due  to random changes in vehicle speed or the environment.  In these cases, robustness of the predetermined temporal order  to such signal timing uncertainties is crucial.

\subsection{Related Work}
\label{sec:rel_work}

Time-critical systems have been studied from various perspectives, and analysis  tools for time-critical systems can  be classified into three categories.  

One category   focuses on the study of real-time systems, see \cite{laplante2004real} and \cite{Jane_liu} for an overview. A particular interest in the study of real-time systems is in scheduling algorithms \cite{sha2004real} that aim at finding an execution order for a set of tasks with corresponding deadlines. Solutions consist, for instance, of periodic scheduling  \cite{serafini1989mathematical} or of event triggered scheduling \cite{tabuada2007event}.  The tools presented in our paper can be seen as complementary to these works as they allow to analyze the risk of  scheduling algorithms.

The second category considers timed automata which allow to analyze real-time systems  \cite{alur1994theory,bengtsson2003timed}. Timed automata enable automatic system verification using model checking tools such as UPPAAL~\cite{behrmann2004tutorial}.  Robustness of timed automata was investigated in \cite{gupta1997robust} and \cite{bendik2021timed}, while control of timed automata was considered in  \cite{maler1995synthesis} and \cite{asarin1998controller}. A connection between the aforementioned scheduling algorithms and timed automata was made in \cite{fersman2006schedulability}. To the best of our knowledge, no one has yet considered the effect of stochastic timing uncertainty and the integration of axiomatic risk theory into a timed automata.

The last category considers formal specification languages such as real-time temporal logics. Metric interval temporal logic (MITL) \cite{alur1996benefits} and signal temporal logic (STL) \cite{maler2004monitoring} enable specifying real-time constraints. It was shown in \cite{alur1996benefits} and \cite{lindemann2019efficient} that checking satisfiability of MITL and STL can be transformed into an emptiness checking problem of a timed automaton. Spatial robustness of MITL specifications over deterministic signals was considered in~\cite{fainekos2009robustness}, while spatial robustness over stochastic signals was analyzed in \cite{bartocci2013robustness} and \cite{bartocci2015system}. Robust linear temporal logic was studied in \cite{anevlavis2021being}. Less attention has been spent on analyzing the temporal robustness of real-time systems subject to temporal logic specifications. Notions of system conformance were introduced in \cite{abbas2014formal,gazda2020logical,deshmukh2015quantifying} to quantify closeness of systems in terms of spatial and temporal closeness of system trajectories. Conformance, however, only allows to reason about temporal robustness with respect to synchronous time shifts of a signal and not with respect to asynchronous time shifts in its sub-signals as captured by our definition of asynchronous temporal robustness. Our notion of asynchronous temporal robustness is especially  important in multi-agent systems where indiviual agents may be delayed and subject to individual timing uncertainties. An orthogonal direction that is worth mentioning is followed by the authors in \cite{penedo2020language} and \cite{kamale2021automata} to find temporal relaxations in control design for time window temporal logic specifications \cite{vasile2017time}. Temporal robustness for STL specifications over deterministic signals was presented in \cite{donze2} and used for control design  in \cite{baras_runtime} and in our recent work \cite{rodionova2021time}.  The notion of asynchronous temporal robustness as presented in this paper is fundamentally different  and
is interpretable in the sense that it quantifies the permissible timing uncertainty of a signal in terms of maximum time shifts in its sub-signals.

The interest in axiomatic risk theory is largely inspired by its longstanding successful use in finance, see e.g., \cite{rockafellar2000optimization}. More recently, risk measures were considered for decision making and control in robotics due to their systematic axiomatization  \cite{majumdar2020should}.  The authors in \cite{jasour2021fast} consider risk assessment in autonomous driving for motion prediction  by deep neural networks, while \cite{novin2021risk} consider risk-aware planning for mobile robots in hospitals. Risk-constrained reinforcement and imitation learning was considered in \cite{chow2017risk} and \cite{lacotte2019risk}, respectively. For signal temporal logic specifications, the use of risk was proposed in our recent work \cite{lindemann2021stl}.  Risk-aware control was further considered in \cite{singh2018framework}, \cite{chapman2019risk}, and \cite{schuurmans2020learning} by using various risk measures, while risk-aware estimation was considered in \cite{kalogerias2020better}.

We conclude with the observation that there is a set of influential works on the analysis and design of real-time systems, but that no one has yet studied the temporal robustness of stochastic signals in this context. Furthermore, no one has yet considered the integration of axiomatic risk theory into the analysis of real-time systems.

\subsection{Contributions and Paper Outline}

Our goal is to analyze the temporal robustness of stochastic signals, and to quantify the risk associated with the signal to not satisfy a specification robustly in time. We make the following contributions.
\begin{itemize}
	\item We define the synchronous and the asynchronous temporal robustness with respect to synchronous and asynchronous time shifts in the sub-signals of a deterministic signal. This definition captures the  permissible  timing uncertainty. 
	\item We define the synchronous and asynchronous temporal robustness risk of a stochastic signal. We  permit the use of various risk measures such as the value-at-risk and show how the temporal robustness risk is estimated from data.
		\item We show that the temporal robustness risk of a stochastic signal under timing uncertainty is upper bounded by the sum of the temporal robustness risk of the nominal stochastic signal and the ``maximum timing uncertainty''.
	\item We extend the previous definitions to signal temporal logic (STL) specifications and define the STL temporal robustness risk. 
	\item We provide empirical evidence and illustrate how the temporal robustness risk can be used in decision making and verification. We also show that, in scenarios where satisfying real-time constraints is of paramount  importance, temporal robustness is advantageous over space robustness notions that are primarily considered in the literature. 
\end{itemize}

 Section \ref{sec:backgound} provides background on stochastic processes and risk measures. In Section \ref{sec:time_robustness}, the synchronous and asynchronous temporal robustness is introduced. The temporal robustness risk is defined in Section \ref{sec:risk}, while the extension to STL specifications is presented in Section \ref{risskk}. Section \ref{sec:estimation} shows how the robustness risk can be estimated from data.  Section \ref{sec:simulations}  presents two case studies on cooperative driving and mobile robots. We conclude  in Section \ref{sec:conclusion}.

\section{Background}
\label{sec:backgound}

 Let $\mathbb{R}$ and $\mathbb{Z}$ be the set of real numbers and integers, respectively, and let $\mathbb{N}$ be the set of natural numbers including zero. Let $\mathbb{R}^n$ be the  $n$-dimensional real vector space. We denote by $\overline{\mathbb{R}}:=\mathbb{R}\cup \{-\infty,\infty\}$ and $\overline{\mathbb{Z}}:=\mathbb{Z}\cup \{-\infty,\infty\}$ the extended real numbers and integers, respectively.  Let $\mathfrak{F}(T,S)$ denote the set of all measurable functions mapping from the domain $T$ into the domain $S$.
 
  In this paper, we aim to define the temporal robustness risk of stochastic  signals. Let us therefore first provide some brief background on stochastic processes and risk measures. We remark  that all  proofs of our technical results can be found in the appendix.

\subsection{Random Variables and Stochastic Processes}
\label{sec:stoch}
Consider  the \emph{probability space} $(\Omega,\mathcal{F},P)$  where $\Omega$ is the sample space, $\mathcal{F}$ is a $\sigma$-algebra of $\Omega$, and $P:\mathcal{F}\to[0,1]$ is a probability measure. Let $Z$ denote a real-valued \emph{random vector}, i.e., a measurable function $Z:\Omega\to\mathbb{R}^n$. When $n=1$, we  say $Z$ is a \emph{random variable}. We refer to $Z(\omega)$ as a realization of $Z$ where $\omega\in\Omega$.   Since $Z$ is a measurable function, a probability space can be defined for $Z$ so that probabilities can be assigned to events related to values of $Z$. Consequently, a cumulative distribution function (CDF) $F_Z(z)$ 
and probability density function (PDF) $f_Z(z)$ 
can be defined for $Z$. 

A \emph{stochastic process} is a function $X:\mathbb{Z}\times \Omega \to \mathbb{R}^n$ where  $X(t,\cdot)$ is a random vector for each fixed time $t\in \mathbb{Z}$. A stochastic process can   be viewed as a collection of random vectors $\{X(t,\cdot)|t\in \mathbb{Z}\}$   defined on a common probability space $(\Omega,\mathcal{F},P)$.  For a fixed $\omega\in\Omega$, the function $X(\cdot,\omega)$ is a \emph{realization} of the stochastic process.  Another equivalent definition is that a stochastic process is a collection of deterministic functions of time  that are indexed by $\Omega$ as $\{X(\cdot,\omega)|\omega\in \Omega\}$.


\subsection{Risk Measures}
\label{sec:risk_measures}

\begin{figure}
	\centering
	\includegraphics[scale=0.5]{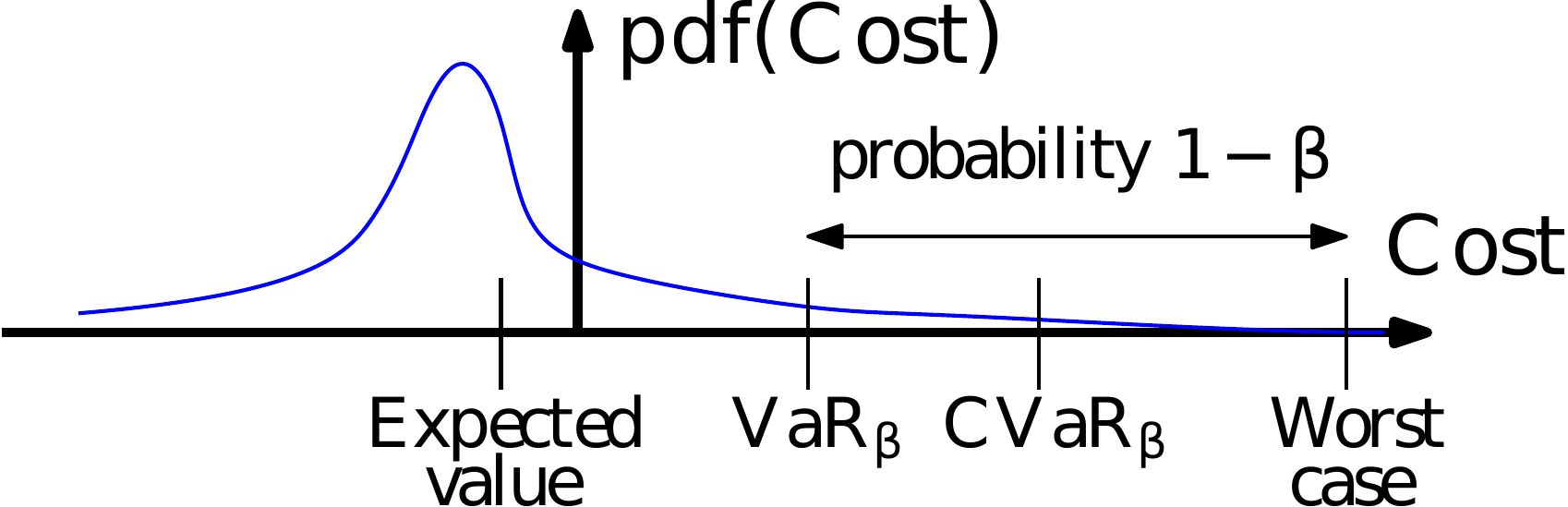}
	\caption{Illustration of various risk measures.}
	\label{ex:2_figure}
\end{figure}

A \emph{risk measure} is a function $R:\mathfrak{F}(\Omega,\mathbb{R})\to \mathbb{R}$ that maps from the set of real-valued random variables to the real numbers. In particular, we refer to the input of a risk measure $R$ as the \emph{cost random variable}. Risk measures  allow for a risk assessment in terms of such cost random variables.  There exists various risk measures, while Figure~\ref{ex:2_figure} particularly illustrates the expected value, the value-at-risk $VaR_\beta$, and the conditional value-at-risk $CVaR_\beta$ at risk level $\beta\in (0,1)$ which are commonly used risk measures. The $VaR_\beta$ of a  random variable $Z\in \mathfrak{F}(\Omega,\mathbb{R})$ is defined as
\begin{align*}
VaR_\beta(Z):=\inf \{ \alpha \in \mathbb{R} |  F_Z(\alpha) \ge \beta \}, 
\end{align*}
i.e., the $1-\beta$  quantile of $Z$. The $CVaR_\beta$ of $Z$ is defined as
\begin{align*}
CVaR_\beta(Z):=\inf_{\alpha \in \mathbb{R}} \; \{\alpha+(1-\beta)^{-1}E([Z-\alpha]^+)\}
\end{align*} 
where $[Z-\alpha]^+:=\max(Z-\alpha,0)$. When the CDF $F_Z$ of $Z$ is continuous, it holds that $CVaR_\beta(Z):=E(Z|Z\ge VaR_\beta(Z))$, i.e., $CVaR_\beta(Z)$ is the expected value of $Z$ conditioned on the events where $Z$ is greater or equal than  $VaR_\beta(Z)$.

Towards an axiomatic risk theory, various properties of risk measures are of interest, see \cite{majumdar2020should} for an overview. While different properties may be desired in certain applications, we particularly use the monotonicity and translational invariance property. 
\begin{itemize}
	\item For two cost random variables $Z,Z'\in \mathfrak{F}(\Omega,\mathbb{R})$, the risk measure $R$ is \emph{monotone} if 
	\begin{align*}
	Z(\omega) \leq Z'(\omega) \text{ for all } \omega\in\Omega \;\; \implies \;\; R(Z) \le R(Z').
	\end{align*}
	\item For a random variable $Z\in \mathfrak{F}(\Omega,\mathbb{R})$, the risk measure $R$ is \emph{translationally invariant} if, for all  $c\in\mathbb{R}$, it holds that
	\begin{align*}
	R(Z + c) = R(Z) + c.
	\end{align*}
\end{itemize} 
Both $VaR_\beta(Z)$ and $CVaR_\beta(Z)$ satisfy these two properties.

\section{Temporal Robustness of Deterministic Signals}
\label{sec:time_robustness}

We  introduce  two  notions  of temporal robustness to quantify how robustly a deterministic signal $x:\mathbb{Z}\to\mathbb{R}^n$ satisfies a real-time constraint with respect to time shifts in $x$.\footnote{We assume here that the time domain $\mathbb{Z}$ of the signal $x$ is the set of integers, i.e., the time domain is unbounded in both directions. This assumption is made without loss of generality and in order to avoid technicalities.} These notions are referred to as the \emph{synchronous temporal robustness} and the \emph{asynchronous temporal robustness}. We  express real-time constraints as constraints $c(x(t),t)\ge 0$ for all $t\in\mathbb{Z}$ where $c:\mathbb{R}^n\times \mathbb{Z}\to \mathbb{R}$ is a measurable constraint function. Later in Section~\ref{risskk}, we consider signal temporal logic (STL) as a  more general means to express complex real-time constraints. The next  example is used throughout this section.
\begin{example}\label{exxx:1}
Consider a signal $x$ that consists of the two sub-signals
\begin{align*}
x_{\{1\}}(t)&:=\sin(a\pi t)\\
x_{\{2\}}(t)&:=-b\sin(1.5a\pi t)
\end{align*}
that are illustrated in red and blue in the left part of Figure \ref{ex:1_example}.\footnote{We choose the non-standard notation of $x_{\{i\}}$ to denote the $i$th sub-signal of $x$ and reserve the notation of $x_i$ for time shifted versions of $x$ as defined in the remainder.} We require these two sub-signals to be $\epsilon$-close within the time interval $[T_1,T_2]$ with parameters $T_1:=145$, $T_2:=155$, $\epsilon:=1$, $a:=0.04$ and $b=1.05$.\footnote{As this paper is concerned with  discrete-time signals, we implicitly assume that a time interval $[T_1,T_2]$ encodes $[T_1,T_2]\cap\mathbb{Z}$.} This requirement is expressed by the constraint function 
\begin{align*}
    c(x(t),t):=\begin{cases}
    1 & \text{for all } t<T_1 \text{ and all } t>T_2\\
    h(x(t)) & \text{for all } t\in [T_1,T_2]
    \end{cases}
\end{align*} 
where $h:\mathbb{R}^2\to \mathbb{R}$ encodes $\epsilon$-closeness of $x_{\{1\}}$ and $x_{\{2\}}$ as
\begin{align*}
h(x(t)):=\epsilon-|x_{\{1\}}(t)-x_{\{2\}}(t)|.
\end{align*}
so that $c(x(t),t)\ge 0$ for all $t\in\mathbb{Z}$ is equivalent to $h(x(t))\ge 0$ for all $t\in [T_1,T_2]$. In the left part of Figure \ref{ex:1_example}, the function $h(x(t))$ is illustrated in black and we can conclude that the requirement is satisfied, i.e., $c(x(t),t)\ge 0$ for all $t\in\mathbb{Z}$.
	\begin{figure*}
	\centering
	\includegraphics[scale=0.6]{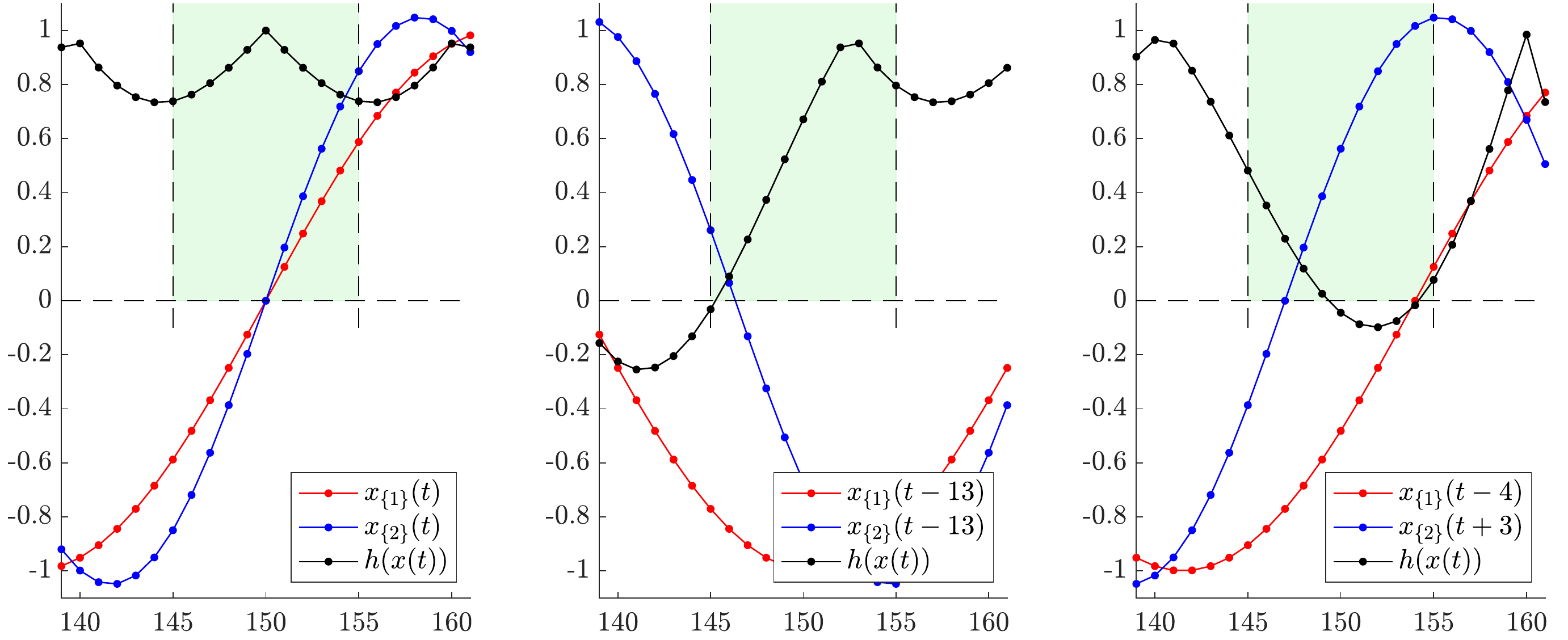}
	\caption{Example \ref{exxx:1} in which the constraint $c$ requires the two signals $x_{\{1\}}(t)$ and $x_{\{2\}}(t)$ to be $\epsilon$-close within the time interval $[145,155]$. Left: The constraint $c$ is satisfied with synchronous temporal robustness $\eta^c(x)=12$ and asynchronous temporal robustness $\theta^c(x)=3$. Middle: The signals $x_{\{1\}}(t)$ and $x_{\{2\}}(t)$ are synchronously shifted by $-13$ so that the constraint $c$ is violated. Right: The signals $x_{\{1\}}(t)$ and $x_{\{2\}}(t)$ are asynchronously shifted by $-4$ and $3$, respectively,  so that the constraint $c$ is violated. }
	\label{ex:1_example}
\end{figure*}
\end{example}

To indicate whether or not a signal $x$ satisfies the constraint $c(x(t),t)\ge 0$ for all $t\in\mathbb{Z}$, let us define the satisfaction function $\beta^c:\mathfrak{F}(\mathbb{Z},\mathbb{R}^n)\to \{-1,1\}$ as
\begin{align*}
    \beta^c(x):=\begin{cases}
    1 & \text{if } \inf_{t\in\mathbb{Z}} c(x(t),t)\ge 0 \\
    -1 & \text{otherwise.}
    \end{cases}
\end{align*}

\subsection{Synchronous Temporal Robustness}

Let us first focus on synchronous time shifts across all sub-signals, i.e., each $x_{\{i\}}(t)$ in $x(t)=(x_{\{1\}}(t),\hdots,x_{\{n\}}(t))$ is shifted in time by the same amount. The \emph{synchronous temporal robustness} is captured by the function $\eta^c:\mathfrak{F}(\mathbb{Z},\mathbb{R}^n)\to \overline{\mathbb{Z}}$ which we define as
\begin{align*}
    \eta^c(x)&:=\beta^c(x)\sup\{\tau\in \mathbb{N}|\forall \kappa\in[-\tau,\tau], \beta^c(x_\kappa)=\beta^c(x)  \}
\end{align*}
where $x_\kappa(t):=x(t+\kappa)$, which is equivalent to $x$ shifted by $\kappa$ time units. Intuitively, $\eta^c(x)$ quantifies the maximum amount by which we can \emph{synchronously} shift $x$ in time while retaining the same behavior as $x$ with respect to the constraint $c$.  Note that $\eta^c(x)$ is non-negative (non-positive) if $\beta^c(x)$ is positive (negative) and that $\eta^c(x)>0$ implies $\beta^c(x)=1$ while $\eta^c(x)<0$ implies $\beta^c(x)=-1$. The following result is a straightforward consequence of the definition of $\eta^c(x)$.   

\begin{corollary}\label{cor:1}
Let $x:\mathbb{Z}\to\mathbb{R}^n$ be a signal and $c:\mathbb{R}^n\times \mathbb{Z}\to\mathbb{R}$ be a constraint function.  For $\xi\in\mathbb{Z}$, it  holds that
\begin{align*}
|\xi|\le |\eta^c(x)| \;\;\; \implies \;\;\; \beta^c(x_\xi)=\beta^c(x).
\end{align*}
\end{corollary}

\begin{example}[continues=exxx:1] 
	Take a look again at Figure~\ref{ex:1_example}.
	The synchronous temporal robustness is $\eta^c(x)=12$, i.e., $x_{\{1\}}$ and $x_{\{2\}}$ can be shifted synchronously by  $12$ time units while still satisfying the constraint $c$. The middle part of Figure \ref{ex:1_example} shows the case where the signals are synchronously shifted  by $-13$ time units, i.e., beyond this limit, so that the constraint $c$ is violated.
\end{example}
We next provide a result that establishes a worst case lower bound for $\eta^c(x_\xi)$ based on $\eta^c(x)$ and $\xi$.  This result is needed later.
\begin{lemma}\label{lem:1}
    Let $x:\mathbb{Z}\to\mathbb{R}^n$ be a signal and $c:\mathbb{R}^n\times \mathbb{Z}\to\mathbb{R}$ be a constraint function. For all $\xi\in\mathbb{Z}$, it  holds that 
    \begin{align*}
        \eta^c(x_\xi)\ge \eta^c(x)-|\xi|.
    \end{align*}
    
    \end{lemma}


\subsection{Asynchronous Temporal Robustness} 
Besides the notion of synchronous temporal robustness, one can think of a notion with asynchronous time shifts across all sub-signals, i.e., each $x_{\{i\}}(t)$ in $x(t)=(x_{\{1\}}(t),\hdots,x_{\{n\}}(t))$ is allowed to be shifted in time by a different amount as opposed to being shifted by the same amount. We capture the \emph{asynchronous temporal robustness} by the function $\theta^c:\mathfrak{F}(\mathbb{Z},\mathbb{R}^n)\to \overline{\mathbb{Z}}$ which we define as
\begin{align*}
		\theta^c(x) &:= \beta^c(x)
		\sup\{\tau\in\mathbb{N}| \forall \kappa_1,\hdots,\kappa_n\in[-\tau,\tau],\beta^c(x_{\bar{\kappa}})=\beta^c(x)\}, 
\end{align*}
where $\bar{\kappa}:=(\kappa_1,\hdots,\kappa_n)$ and
 \begin{align*}
x_{\bar{\kappa}}(t):=(x_{\{1\}}(t+\kappa_1),\hdots,x_{\{n\}}(t+\kappa_n)).
\end{align*}
The interpretation of $\theta^c(x)$ is that it quantifies the amount by which we can \emph{asynchronously shift} $x$ in time while retaining the same behavior as $x$ with respect to the constraint $c$. The next corollary is  a consequence of this definition and we particularly use the notation $\bar{\xi}:=(\xi_1,\hdots,\xi_n)$ for $\xi_1,\hdots,\xi_n\in \mathbb{Z}$.

\begin{corollary}\label{cor:2}
Let $x:\mathbb{Z}\to\mathbb{R}^n$ be a signal and $c:\mathbb{R}^n\times \mathbb{Z}\to\mathbb{R}$ be a constraint function.  For $\xi_1,\hdots,\xi_n\in \mathbb{Z}$, it holds that
\begin{align*}
\max(|\xi_1|,\hdots,|\xi_n|)\le |\theta^c(x)| \;\;\; \implies \;\;\; \beta^c(x_{\bar{\xi}})=\beta^c(x).
\end{align*} 
\end{corollary}

\begin{example}[continues=exxx:1] 
	Recall again Figure \ref{ex:1_example}. The asynchronous temporal robustness is $\theta^c(x)=3$, i.e., $x_{\{1\}}$ and $x_{\{2\}}$ can be shifted asynchronously by $3$ time units while still satisfying the constraint $c$. The right part of Figure \ref{ex:1_example} shows the case where the signals are shifted by $-4$ and $3$, respectively, so that the constraint $c$ is violated.
\end{example}

We next state a result that is similar to Lemma \ref{lem:1} but for the asynchronous temporal robustness. It  establishes a worst case lower bound for $\theta^c(x_{\bar{\xi}})$ based on $\theta^c(x)$ and $\bar{\xi}$. 
\begin{lemma}\label{lem:2}
    Let $x:\mathbb{Z}\to\mathbb{R}^n$ be a signal and $c:\mathbb{R}^n\times \mathbb{Z}\to\mathbb{R}$ be a constraint function. For all $\xi_1,\hdots,\xi_n\in \mathbb{Z}$, it holds that 
    \begin{align*}
        \theta^c(x_{\bar{\xi}})\ge \theta^c(x)-\max(|\xi_1|,\hdots,|\xi_n|).
    \end{align*}
    
\end{lemma}

Some remarks are in place regarding the presented notions and computational aspects.

\begin{remark}\label{reemmmm}
	To calculate $\eta^c(x)$ and $\theta^c(x)$ in practice, we let $r\in\mathbb{N}$ be a bound on the maximum temporal robustness (in absolute value) that we are interested in. This allows us to replace the statement $\tau\in\mathbb{N}$ in the definitions of $\eta^c(x)$ and $\theta^c(x)$ by the statement $\tau\in[0,r]$.  The complexity of calculating $\eta^c(x)$, measured in the number of shifted versions of $x$ that one may need to  construct, scales polynomially with $r$ and is  $O(r)$. In particular, at most $2r+1$ shifted versions of $x$ need to be constructed. On the other hand, the complexity of $\theta^c(x)$ scales polynomially in $r$ and exponentially in the state dimension $n$ and is $O(r^n)$, i.e., at most $(2r+1)^n$ shifted versions of $x$ need to be  constructed. 
\end{remark}

\begin{remark}\label{rem:comp_compl}
	There are various ways to reduce the  complexity of calculating $\theta^c(x)$. First, observe that subsets of states  of dynamical systems, e.g., position and velocity of a vehicle, are often subject to the same timing uncertainty. Under this assumption, we can group states in these subsets together and shift them  synchronously which simplifies the calculation of $\theta^c(x)$. More explanation is provided  Section~\ref{sec:simulations}. Second, a natural way to reduce the complexity in practice is to downsample the signal $x$. Third, depending on the coupling of states in $c(x(t),t)$ one may be  able to decompose the calculation of $\theta^c(x)$ into smaller sub-problems that can be solved in parallel.
\end{remark}

\begin{remark}\label{rem:time_scaling}
	These notions of temporal robustness focus on shifts of $x$ in time, while one may also be interested in time scaling effects, e.g., $x(\tau \cdot t)$ for some $\tau\in\mathbb{N}$, which we are not considering in this paper.
\end{remark}

We conclude this section by comparing the two presented definitions. The asynchronous temporal robustness $\theta^c(x)$ is sensitive to asynchronous time shifts and hence more general than the synchronous temporal robustness $\eta^c(x)$.  The  synchronous temporal robustness, however,  is easier to compute as argued in Remark~\ref{reemmmm} and provides an upper bound to $\theta^c(x)$ as shown next.
\begin{corollary}\label{cor:eta_theta}
	Let $x:\mathbb{Z}\to\mathbb{R}^n$ be a signal and $c:\mathbb{R}^n\times \mathbb{Z}\to\mathbb{R}$ be a constraint function. Then it holds that  
	\begin{align*}
	|\theta^c(x)| \leq |\eta^c(x)|.
	\end{align*} 
\end{corollary}

\section{Temporal Robustness of Stochastic Signals}
\label{sec:risk}

So far, we  evaluated the constraint function $c(x(t),t)$ over  signals $x:\mathbb{Z}\to\mathbb{R}^n$. We are now interested in evaluating $c(X(t,\cdot),t)$ over a stochastic process $X$. As $c$ is a measurable function and $X(t,\cdot)$ is a random variable, it follows that $c(X(t,\cdot),t)$ is a random variable. We next show that 
 $\eta^c(X)$ and $\theta^c(X)$ are also random variables. 
\begin{theorem}\label{thm:1}
    Let $X:\mathbb{Z}\times\Omega\to\mathbb{R}^n$ be a discrete-time stochastic process and  $c:\mathbb{R}^n\times \mathbb{Z}\to \mathbb{R}$ be a measurable constraint function. Then $\eta^c(X(\cdot,\omega))$ and $\theta^c(X(\cdot,\omega))$ are measurable in $\omega$, i.e.,  $\eta^c(X)$ and $\theta^c(X)$ are random variables.
\end{theorem}

In general, it is desirable if the distributions of $\eta^c(X)$ and $\theta^c(X)$  are concentrated around larger values to be robust with respect to time shifts in $X$. This is why we view $-\eta^c(X)$ and $-\theta^c(X)$ as cost random variables. Following this interpretation, we define the \emph{synchronous temporal robustness risk} and the \emph{asynchronous temporal robustness risk} by applying risk measures $R$ to $-\eta^c(X)$ and $-\theta^c(X)$.
\begin{definition}[Synchronous and Asynchronous Temporal Robustness Risk]\label{tr_uniform}
Let $X:\mathbb{Z}\times\Omega\to\mathbb{R}^n$ be a discrete-time stochastic process and  $c:\mathbb{R}^n\times \mathbb{Z}\to \mathbb{R}$ be a measurable constraint function. The synchronous temporal robustness risk is defined as
 	\begin{align*}
 	R(-\eta^c(X))
 	\end{align*}  
while the asynchronous temporal robustness risk is defined as
 	\begin{align*}
 	R(-\theta^c(X)).
 	\end{align*}  
\end{definition}

The previous definitions quantify the risk of the stochastic process $X$ to not  satisfy the   constraint $c$ robustly  with respect to synchronous and asynchronous time shifts in $X$. Note particularly that  $R(-\eta^c(X))<0$ and $R(-\theta^c(X))<0$ can be given a sound interpretation, e.g., for the value-at-risk $VaR_\beta$  at level $\beta$ (the $1-\beta$ worst case quantile) the interpretation is that at most $1-\beta$ percent of the realizations of $X$ violate the constraint $c$. 

We next analyze the  properties of the synchronous and the asynchronous temporal robustness risks. For  $d\in\mathbb{N}$ and a shift random variable $\xi\in\mathfrak{F}(\Omega,[-d,d])$, let us now define the time shifted version $X_{\xi}$ of $X$ in analogy to $x_\xi$ from the previous section as
\begin{align*}
X_{\xi}(t,\omega):=X(t+\xi(\omega),\omega).
\end{align*} 
In other words, $X_{\xi}$ is equivalent to the stochastic process $X$ synchronously shifted in time by the shift random variable $\xi$. The meaning of the synchronous temporal robustness risk with respect to the shift random variable $\xi$ is   shown in the next result.
\begin{theorem}\label{thm:2}
	Let $X:\mathbb{Z}\times\Omega\to\mathbb{R}^n$ be a discrete-time stochastic process and  $c:\mathbb{R}^n\times \mathbb{Z}\to \mathbb{R}$ be a measurable constraint function. Let the risk measure $R$ be monotone and translationally invariant. For  $d\in\mathbb{N}$ and a shift random variable $\xi\in\mathfrak{F}(\Omega,[-d,d])$, it holds that
	\begin{align*}
	R(-\eta^c(X_\xi))\le R(-\eta^c(X))+d.
	\end{align*}
\end{theorem}

The synchronous temporal robustness risk $R(-\eta^c(X_\xi))$ of  $X_\xi$ can hence not exceed the sum of the synchronous robustness risk $R(-\eta^c(X))$ of $X$ and the bound $d$ of the  shift random variable $\xi$. Importantly, Theorem~\ref{thm:2} tells us that
\begin{align*}
R(-\eta^c(X))< 0 \;\;\; \implies \;\;\; R(-\eta^c(X_\xi))< 0
\end{align*}
for all $\xi\in\mathfrak{F}(\Omega,[-d,d])$ where $d$ is such that $d< |R(-\eta^c(X))|$.  The next corollary specializes Theorem \ref{thm:2} to deterministic $\xi$.
\begin{corollary}
	Under the  conditions of Theorem \ref{thm:2} and for  $\xi\in\mathbb{Z}$,  it holds that 
	\begin{align*}
	R(-\eta^c(X_\xi))\le R(-\eta^c(X))+|\xi| .
	\end{align*}
\end{corollary}

Similar results and interpretations can be obtained for the asynchronous robustness risk. For $d\in\mathbb{N}$ and shift random variables $\xi_1,\hdots,\xi_n\in\mathfrak{F}(\Omega,[-d,d])$, define next $X_{\bar{\xi}}$ in analogy to $x_{\bar{\xi}}$ from the previous section as
\begin{align*}
X_{\bar{\xi}}(t,\omega):=(X_{\{1\}}(t+\xi_1(\omega),\omega),\hdots, X_{\{n\}}(t+\xi_n(\omega),\omega)).
\end{align*}
 We can now present the analogous result of Theorem \ref{thm:2}. 

\begin{theorem}\label{thm:3}
	Let $X:\mathbb{Z}\times\Omega\to\mathbb{R}^n$ be a discrete-time stochastic process and  $c:\mathbb{R}^n\times \mathbb{Z}\to \mathbb{R}$ be a measurable constraint function. Let the risk measure $R$ be monotone and  translationally invariant. For $d\in\mathbb{N}$ and shift random variables $\xi_1,\hdots,\xi_n\in\mathfrak{F}(\Omega,[-d,d])$, it holds that
	\begin{align*}
	 R(-\theta^c(X_{\bar{\xi}}))\le R(-\theta^c(X))+d.
	\end{align*}
\end{theorem}

The asynchronous temporal robustness risk $R(-\theta^c(X_{\bar{\xi}}))$ of  $X_{\bar{\xi}}$ can hence not exceed the sum of the asynchronous robustness risk $R(-\theta^c(X))$ of $X$ and the bound $d$ of the  shift random variables ${\bar{\xi}}$. Importantly, Theorem \ref{thm:3} now tells us  that
\begin{align*}
R(-\theta^c(X))<0 \;\;\; \implies \;\;\;  R(-\theta^c(X_{\bar{\xi}}))<0
\end{align*}
for all $\xi_1,\hdots,\xi_n\in\mathfrak{F}(\Omega,[-d,d])$ where $d$ is such that $d< |R(-\theta^c(X))|$. The next corollary specializes Theorem \ref{thm:3} to deterministic $\xi_1,\hdots,\xi_n$.

\begin{corollary}\label{corrrrrrr}
	Under the  conditions of Theorem \ref{thm:3} and for $\xi_1,\hdots,\xi_n\in\mathbb{Z}$,  it holds that 
	\begin{align*}
	R(-\theta^c(X_{\bar{\xi}}))\le R(-\theta^c(X))+\max(|\xi_1|,\hdots,|\xi_n|).
	\end{align*}
\end{corollary}

\section{The STL Temporal Robustness Risk}
\label{risskk}

We now extend the notion of synchronous and asynchronous temporal robustness to general specifications formulated in signal temporal logic (STL) introduced in \cite{maler2004monitoring}. STL specifications are  constructed from predicates  $\mu:\mathbb{R}^n\to\{-1,1\}$. Typically, these predicates are defined via predicate functions $h:\mathbb{R}^n\to\mathbb{R}$ as
\begin{align*}
\mu(\zeta):=\begin{cases}
1 & \text{if } h(\zeta)\ge 0\\
-1&\text{otherwise}
\end{cases}
\end{align*}
for $\zeta\in\mathbb{R}^n$. The syntax of STL is then recursively defined as 
\begin{align*}
\phi \; ::= \; \top \; | \; \mu \; | \;  \neg \phi \; | \; \phi' \wedge \phi'' \; | \; \phi'  U_I \phi'' \; | \; \phi' \underline{U}_I \phi'' \,
\end{align*}
where $\top$ is the logical true element, $\phi'$ and $\phi''$ are STL formulas and where $U_I$ is the future until operator with $I\subseteq \mathbb{R}_{\ge 0}$, while $\underline{U}_I$ is the past until-operator. The operators $\neg$ and $\wedge$ encode negation and conjunction. One can further define the operators $\phi' \vee \phi'':=\neg(\neg\phi' \wedge \neg\phi'')$ (disjunction),
$F_I\phi:=\top U_I \phi$ (future eventually),
$\underline{F}_I\phi:=\top \underline{U}_I \phi$ (past eventually),
$G_I\phi:=\neg F_I\neg \phi$ (future always),
$\underline{G}_I\phi:=\neg \underline{F}_I\neg \phi$ (past always).

\emph{Semantics.} We now define the satisfaction function $\beta^\phi:\mathfrak{F}(\mathbb{Z},\mathbb{R}^n)\times \mathbb{Z} \to \{-1,1\}$. In particular, $\beta^\phi(x,t)=1$ indicates that the signal $x$ satisfies the formula $\phi$ at time $t$, while $\beta^\phi(x,t)=-1$ indicates that $x$ does not satisfy $\phi$ at time $t$. Following \cite{maler2004monitoring}, the semantics $\beta^\phi(x,t)$ of an STL formula $\phi$ are inductively defined as
	\begin{align*}
	\beta^\top(x,t)&:=1,  \\
	\beta^\mu(x,t)&:=\begin{cases}
	1 &\text{ if }	h(x(t))\ge 0 \\	
	-1 &\text{ otherwise, }	
	\end{cases}\\
	\beta^{\neg\phi}(x,t)&:= \neg \beta^{\phi}(x,t),\\
	\beta^{\phi' \wedge \phi''}(x,t)&:=\min(\beta^{\phi'}(x,t),\beta^{\phi''}(x,t)),\\
	\beta^{\phi' U_I \phi''}(x,t)&:=\sup_{t''\in (t\oplus I)\cap\mathbb{Z}}\Big( \min\big(\beta^{\phi''}(x,t''),\inf_{t'\in[t,t'']}\beta^{\phi'}(x,t')\big)\Big),\\
	\beta^{\phi' \underline{U}_I \phi''}(x,t)&:=\sup_{t''\in (t\ominus I)\cap\mathbb{Z}}\Big( \min\big(\beta^{\phi''}(x,t''),\inf_{t'\in[t'',t]}\beta^{\phi'}(x,t')\big)\Big)
	\end{align*}
where $\oplus$ and $\ominus$ denote the Minkowski sum operator and the Minkowski difference operator, respectively.

\emph{STL Temporal Robustness.} Besides calculating $\beta^\phi(x,t)$ to determine satisfaction of $\phi$ by $x$ at time $t$, one can again calculate how robustly $x$ satisfies $\phi$ at time $t$ with respect to synchronous and asynchronous  time shifts in $x$.  The \emph{synchronous STL temporal robustness} at time $t$ is
\begin{align*}
    \eta^\phi(x,t)&:=\beta^\phi(x,t)\sup\{\tau\in \mathbb{N}|\forall \kappa\in[-\tau,\tau], \beta^\phi(x_\kappa,t)=\beta^\phi(x,t)  \}
\end{align*}
while the \emph{asynchronous STL temporal robustness} at time $t$ is
\begin{align*}
		\theta^\phi(x,t) := \beta^\phi(x,t)
		\sup\{\tau\in\mathbb{N}| \forall \kappa_1,\hdots,\kappa_n\in[-\tau,\tau],\beta^\phi(x_{\bar{\kappa}},t)=\beta^\phi(x,t)\}. 
\end{align*}

For a fixed $t$, one can easily show that the same results presented for $\eta^c(x)$ and $\theta^c(x)$  in Corollaries \ref{cor:1} and \ref{cor:2}  also hold for $\eta^\phi(x,t)$ and $\theta^\phi(x,t)$, respectively. 

\begin{corollary}\label{cor:4}
	Let $x:\mathbb{Z}\to\mathbb{R}^n$ be a signal and $\phi$ be an STL formula. For $\xi\in\mathbb{Z}$, it holds that
	\begin{align*}
	|\xi|\le |\eta^\phi(x,t)| \;\;\; \implies \;\;\; \beta^\phi(x_\xi,t)=\beta^\phi(x,t).
	\end{align*}
	Furthermore, for $\xi_1,\hdots,\xi_n\in \mathbb{Z}$ it holds that
	\begin{align*}
	\max(|\xi_1|,\hdots,|\xi_n|)\le |\theta^\phi(x,t)| \;\;\; \implies \;\;\; \beta^\phi(x_{\bar{\xi}},t)=\beta^\phi(x,t).
	\end{align*} 
\end{corollary}

\begin{remark}
	The synchronous and asynchronous STL temporal robustness differ from the temporal robustness defined in \cite{donze2}. We consider  time shifts in $x$ over $\phi$ via $\beta^\phi(x,t)$ instead of time shifts in $x$ over each predicate $\mu$ in $\phi$ via $\beta^\mu(x,t)$. This enables Corollary \ref{cor:4} which does not hold for the definition in \cite{donze2}.
\end{remark}

\emph{STL Temporal Robustness Risk.} We now interpret the STL formula $\phi$ over the stochastic process $X$ instead of a deterministic signal $x$. In  \cite[Theorem 1]{lindemann2021stl}, it was shown that $\beta^\phi(X,t)$ is a random variable. As in Theorem  \ref{thm:1}, it can hence be shown that $\eta^\phi(X,t)$ and $\theta^\phi(X,t)$ are random variables. 
\begin{definition}[Synchronous and Asynchronous  STL Temporal Robustness Risk]
 	 Let $X:\mathbb{Z}\times\Omega\to\mathbb{R}^n$ be a discrete-time stochastic process and $\phi$ be an STL formula. The synchronous STL temporal robustness risk is defined as 
 	\begin{align*}
 	R(-\eta^{\phi}(X,t))
 	\end{align*}  
	while the asynchronous STL temporal robustness risk is defined as
 	\begin{align*}
 	R(-\theta^{\phi}(X,t)).
 	\end{align*}  
\end{definition}

The following result can be derived similarly to the results in Sections \ref{sec:risk} and is stated without a proof.

\begin{theorem}\label{thm:4}
	Let $X:\mathbb{Z}\times\Omega\to\mathbb{R}^n$ be a discrete-time stochastic process   and $\phi$ be an STL formula. Let the risk measure $R$ be monotone and  translationally invariant. For  $d\in\mathbb{N}$ and a shift random variable $\xi\in\mathfrak{F}(\Omega,[-d,d])$, it holds that 
	\begin{align*}
	R(-\eta^\phi(X_\xi,t))\le R(-\eta^\phi(X,t))+d.
	\end{align*}
	For shift random variables $\xi_1,\hdots,\xi_n\in\mathfrak{F}(\Omega,[-d,d])$, it holds that
	\begin{align*}
	R(-\theta^\phi(X_{\bar{\xi}},t))\le R(-\theta^\phi(X,t))+d.
	\end{align*}

\end{theorem}

\section{Estimation of the Temporal Robustness Risk from Data}
\label{sec:estimation}

Motivated by recent interest in data-driven verification, we show how the temporal robustness risk $R(-\eta^c(X))$ and $R(-\theta^c(X))$ as well as the STL temporal robustness risk $R(-\eta^\phi(X,t))$ and $R(-\theta^\phi(X,t))$ can be estimated from data. Let us, for convenience,  define a random variable $Z$ that corresponds to
\begin{align*}
Z\in \{-\eta^c(X), -\theta^c(X),-\eta^\phi(X,t),-\theta^\phi(X,t)\}
\end{align*}
 depending on the case of interest. For further convenience, let us define the tuple
\begin{align*}
\mathcal{Z}:=(Z^1,\hdots,Z^N)
\end{align*}
where again depending on the case of interest  
\begin{align*}
Z^i\in\{-\eta^c(X^i),-\theta^c(X^i),-\eta^\phi(X^i,t),-\theta^\phi(X^i,t)\}
\end{align*}
 and where $X^1,\hdots,X^N$ are $N$ observed realizations of $X$. Without loss of generality, assume that the tuple $\mathcal{Z}$ is sorted in increasing order, i.e., $Z^i\le Z^{i+1}$ for each $i\in\{1,\hdots,N-1\}$.

We consider the value-at-risk (VaR) as a risk measure. For a risk level of $\beta\in(0,1)$, recall that
$VaR_\beta(Z):= \inf\{\alpha\in\mathbb{R}|F_{Z}(\alpha)\ge \beta\}$. Note now that the distribution function $F_Z$ is discontinuous as we consider  discrete-time stochastic processes $X$. To estimate $VaR_\beta(Z)$, we  use \cite[Lemma 3]{nikolakakis2021quantile} to be able to deal with discontinuous distribution functions $F_Z$ using order statistics.  The following result follows from \cite[Lemma 3]{nikolakakis2021quantile} and is stated without proof.
\begin{proposition}\label{prop:1}
	Let $\delta\in(0,1)$ be a failure probability, $\beta\in(0,1)$ be a risk level, and
		$\gamma:=\sqrt{{\ln(2/\delta)}/{2N}}$
		be such that $\gamma\le \min \{\beta,1-\beta\}$.	Then,  with probability of at least $1-\delta$, it holds that
		\begin{align*}
		\underline{VaR}_\beta(\mathcal{Z})\le VaR_\beta(Z)\le \overline{VaR}_\beta(\mathcal{Z})
		\end{align*} 
		where the upper and lower bounds are defined as\footnote{We let $\lceil N(\beta+\gamma)\rceil$ and $\lfloor N(\beta-\gamma)\rfloor$ be the nearest integer to $N(\beta+\gamma)$  from above and the nearest integer to $N(\beta-\gamma)$ from below, respectively. }
	\begin{align*}
	\overline{VaR}_\beta(\mathcal{Z})&:=Z^{\lceil N(\beta+\gamma)\rceil}\\
	\underline{VaR}_\beta(\mathcal{Z})&:=Z^{\lfloor N(\beta-\gamma)\rfloor}.
	\end{align*}
\end{proposition}

In case that $Z$ is a random variable with bounded support, we can derive similar estimators as in Proposition \ref{prop:1} for the conditional value-at-risk, see e.g., \cite{wang2010deviation}, which is omitted here for brevity. 

\section{Case Studies}
\label{sec:simulations}

We present two case studies in this section. The first case study considers a T-intersection in a cooperative driving scenario with a constraint function $c$ encoding collision avoidance at the intersection. The second case study considers two mobile robots under an STL specification $\phi$ encoding servicing tasks. Our code is available at \href{https://github.com/temporalrobrisk/Temporal-Robustness-Risk/}{https://tinyurl.com/temporalrob} and animations of the two case studies can be found at \href{https://www.youtube.com/channel/UCcypQ0_IYt02cBXv9nSNpYA/videos}{https://tinyurl.com/temporalsim}.

\subsection{Cooperative Driving at a T-Intersection}

This example is inspired by the  Grand Cooperative Driving Challenge that aimed at communication-enabled cooperative autonomous driving \cite{englund2016grand}. We consider a T-intersection scenario  where timing uncertainty, e.g., resulting from communication delays or communication errors  caused by malfunctioning communication protocols,  may lead to a violation of the specification.

Particularly, consider three cars at a T-intersection as illustrated in Figure \ref{ex:1_figure}. At time $t=0$, each car is supposed to transmit their estimated positions. The green vehicle is non-cooperative, e.g., a human driver, while the blue and red vehicles are cooperative in the sense that they adjust their speeds based on the transmitted information to avoid collisions. For  an intersection of size $\epsilon_1:=10$  and a minimum safety distance of $\epsilon_2:=15$, the safety specification is based upon
\begin{align*}
h_1(x)&:=\epsilon_1-\|x_\text{green}\|\\
h_2(x)&:=\|x_\text{green}-x_\text{red}\|-\epsilon_2\\
h_3(x)&:=\|x_\text{green}-x_\text{blue}\|-\epsilon_2.
\end{align*}
When $h_1(x)\ge 0$, the green car is within $\epsilon_1$ radius of the intersection, while $h_2(x)\ge 0$ and $h_3(x)\ge 0$ indicate that a minimum safety distance of  $\epsilon_2$ is violated between cars. We consider the Boolean specification
\begin{align*}
(h_1(x)\ge 0) \implies [(h_2(x)\ge 0)\wedge (h_3(x)\ge 0)]
\end{align*}
which is expressed by the constraint function
\begin{align*}
c(x,t):=\max(-h_1(x),\min(h_2(x),h_3(x))).
\end{align*}
Let us assume  initial positions of $x_\text{green}(0):=(-5,300)$, $x_\text{red}(0):=(300,5)$, and $x_\text{blue}(0):=(-300,-5)$. We have two scenarios $S_1$ and $S_2$. In both $S_1$ and $S_2$, the non-collaborative car is driving with velocity $v_\text{green}:=15$. In $S_1$, let $v_\text{red}:=12$ and $v_\text{blue}:=18$ so that first the blue, then the green, and then the red car pass the intersection. In $S_2$, let $v_\text{red}:=18$ and $v_\text{blue}:=12$ so that first the red, then the green, and the blue car pass the intersection. Note that we have chosen a simulation horizon of $60$ time units and a time discretization of $0.1$.\footnote{To obtain signals that are defined on $\mathbb{Z}$, we extend all simulated trajectories that are defined on the interval $[0,60]$  to the left and right with the corresponding left and right endpoints of the trajectory.} In both scenarios, we have $\beta^c(x)=1$, i.e., the specification is satisfied, and $\eta^c(x)=+\infty$ for the synchronous temporal robustness. The asynchronous temporal robustness, however, is $\theta^c(x)=10$ in $S_1$ and $\theta^c(x)=8$ in $S_2$. This indicates that scenario $S_1$ is preferable from a temporal robustness point of view. Interestingly, the spatial robustness, here defined as $\inf_{t\in \mathbb{Z}} c(x(t),t)$, is $8.5$ in $S_1$ and $16.87$ in $S_2$ indicating that $S_2$ is preferable. This example highlights the difference between temporal and spatial robustness.

	\begin{figure*}
	\centering
	\includegraphics[scale=0.6]{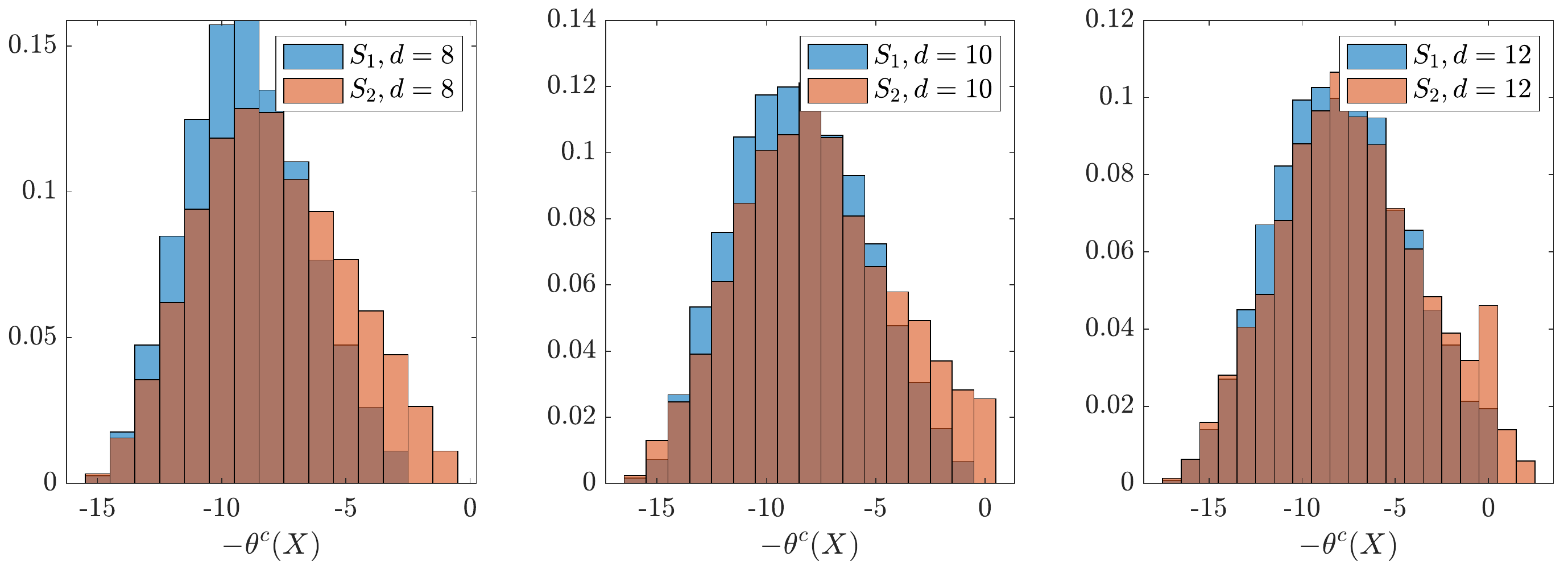}
	\caption{Histogram of $-\theta^c(X)$ in the T-intersection scenario in case of uncertainty in the communication between cars.}
	\label{ex:3_figure}
\end{figure*}

\subsubsection{Communication Delays}
Let us first model communication delays and errors. Let therefore the initial time of each car be  random, i.e., let $x_\text{green}(t_0^\text{g}):=(-5,300)$, $x_\text{red}(t_0^\text{r}):=(300,5)$, and $x_\text{blue}(t_0^\text{b}):=(-300,-5)$ where $t_0^\text{r}$, $t_0^\text{g}$, and $t_0^\text{b}$ are uniformly distributed within the interval $[-d,d]\subseteq \mathbb{Z}$. We consider the cases $d\in\{0,8,10,12\}$, set $\delta:=0.01$ and $Z:=-\theta^c(X)$, and collected $N:=10000$ realizations of $X$. We additionally calculate estimates of the conditional value-at-risk based on \cite{rockafellar2000optimization}, denoted by  $\widehat{CVaR}_{\beta}(\mathcal{Z})$.\footnote{Note that $\widehat{E}$ and $\widehat{CVaR}$ are estimates and not an upper bound like $\overline{VaR}$.} The numerical results for these risk measures are shown in Table \ref{table:1}, while Figure \ref{ex:3_figure} shows the histograms of $-\theta^c(X)$.  We  make the following observations:
\begin{itemize}
	\item As lower values of $R(-\theta^c(X))$ imply less risk, scenario $S_1$ is preferable across all considered risk measures. This illustrates how the temporal robustness risk can be used to make risk-aware decisions.
	\item With $S_1$ and $d=0$, we have that $R(-\theta^c(X))=-10$. By Theorem \ref{thm:3}, we expect $R(-\theta^c(X))\le 0$ for $S_1$ and $d=10$. Indeed, this is shown empirically in the table below. Similarly, as $R(-\theta^c(X))=-8$ for $S_2$ and $d=0$, it holds that $R(-\theta^c(X))\le 0$ for $S_2$ and $d=8$.
	\item The upper and lower bounds $\overline{VaR}_\beta$ and $\underline{VaR}_\beta$ are tight and provide a good approximation of ${VaR}_\beta$. 
	\item The last column shows $\#:=\sum_{i=1}^N \mathbb{I}(\beta^c(X^i)=-1)$, i.e., the number of realizations that violate $c$. Note that the numbers are in line with Corollary \ref{cor:2} and the interpretation of  VaR.
\end{itemize}
\begin{table}[h!]
	\centering
	\begin{tabular}{|p{2cm}|p{1.25cm}|p{1.25cm}|p{1.25cm}|p{1.25cm}|p{1.5cm}|p{0.6cm}|}
		\hline
		\backslashbox{}{Risk} & $\overline{VaR}_{0.95}$ & $\underline{VaR}_{0.95}$  & $\overline{VaR}_{0.98}$ &  $\underline{VaR}_{0.98}$ &  $\widehat{CVaR}_{0.95}$ & $\#$\\
		\hline
		$S_1$, $d=0$ & -10 & -10 & -10 & -10 & -10 & 0\\
		$S_1$, $d=8$ & -4  & -5 & -3 & -4 & -4  & 0\\
		$S_1$, $d=10$ & -3 & -4 & -1 & -3 & -2.4  & 0\\
		$S_1$, $d=12$ & -1  & -2 &  0 & -1 & -0.8 & 57 \\
		\hline
		$S_2$, $d=0$ & -8 & -8 & -8  & -8 & - 8 & 0 \\
		$S_2$, $d=8$  & -2 & -3 &  -1 & -2 & -2  & 0 \\
		$S_2$, $d=10$  & -1& -2 & 0 & -1 & -0.5 & 78 \\
		$S_2$, $d=12$ & 0  & -1 & 2  & 0 & 0.5 & 404 \\
		\hline
	\end{tabular}
	\caption{Risk estimates of $-\theta^c(X)$ where the initial  time of each car is  uniformly distributed within $[-d,d]$.}
	\label{table:1}
\end{table}

	\subsubsection{Communication Delays and Control Disturbances} Usually, the cruise controller of a car is not exact. We model such errors by adding Gaussian distributed noise to $v_\text{green}$, $v_\text{red}$, and $v_\text{blue}$. For $Z:=-\theta^c(X)$, $N:=10000$, and $\delta:=0.01$, the results for various risk measures and $-\theta^c(X)$ are shown in Table \ref{table:2}.
\begin{table}[h!]	
	\centering
\begin{tabular}{|p{2cm}|p{1.25cm}|p{1.25cm}|p{1.25cm}|p{1.25cm}|p{1.25cm}|p{0.6cm}|}
	\hline
	\backslashbox{}{Risk} & $\overline{VaR}_{0.85}$ & $\overline{VaR}_{0.9}$ & $\overline{VaR}_{0.95}$ & $\overline{VaR}_{0.98}$  & $\#$\\
	\hline
	$S_1$ & -6 & -5 & -3 & 0  & 20\\
	$S_2$ & -5 & -5 & -4 & -2 &  1 \\
	\hline
\end{tabular}
\caption{Risk estimates of $-\theta^c(X)$ where the velocity of each car is affected by additive Gaussian noise with variance $0.3^2$.}
\label{table:2}
\end{table}

\begin{figure}
	\centering
	\includegraphics[scale=0.4]{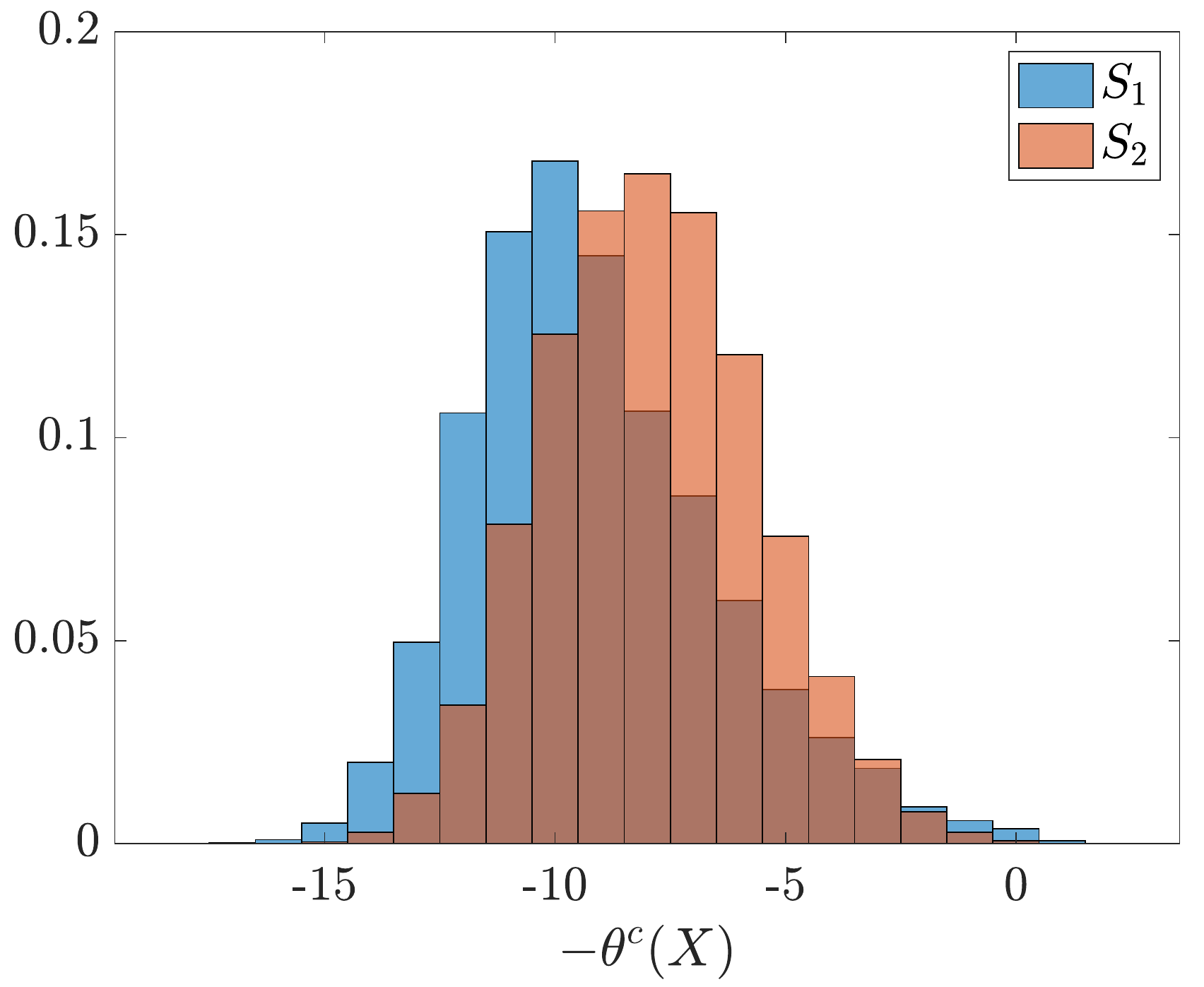}
	\caption{Histogram of $-\theta^c(X)$ in the T-intersection scenario in case of uncertainty in the cruise controllers.}
	\label{ex:4_figure}
\end{figure}

Additionally, Figure \ref{ex:4_figure} shows the histograms of $-\theta^c(X)$. Interestingly, the distribution of $S_2$ is Gaussian shaped, while the distribution of $S_1$ has a long tail of ``bad'' events. This is reflected in the calculated risks, i.e., for $\overline{VaR}_{0.85}$ and $\overline{VaR}_{0.9}$ the $S_1$ scenario is preferable. However, $\overline{VaR}_{0.95}$ and $\overline{VaR}_{0.98}$ detect the non-safe tail events in $S_1$ (the blue tail in Figure \ref{ex:4_figure}) and label $S_2$ as preferable. This is confirmed by the number of realizations that violate $c$ in the rightmost column of the table and denoted by $\#$. 

So far, only the velocities were affected by noise in this subsection. To utilize the obtained information in terms of Corollary \ref{corrrrrrr}, we consider the starting times  $t_0^\text{r}:=-5$, $t_0^\text{g}:=5$ , and $t_0^\text{b}:=2$. The results for $-\theta^c(X)$, aligning with Corollary \ref{corrrrrrr}, are shown in Table \ref{table:3}.
\begin{table}[h!]
	\centering
\begin{tabular}{|p{2cm}|p{1.25cm}|p{1.25cm}|p{1.25cm}|p{1.25cm}|p{0.6cm}|}
	\hline
	\backslashbox{}{Risk} & $\overline{VaR}_{0.85}$ & $\overline{VaR}_{0.9}$ & $\overline{VaR}_{0.95}$ & $\overline{VaR}_{0.98}$   & $\#$\\
	\hline
	$S_1$ & -1 & 0 & 1 & 4 & 604\\
	$S_2$ & -9 & -8 & -7 & -5 &  1 \\
	\hline
\end{tabular}
\caption{Risk estimates of $-\theta^c(X)$ where the velocity of each car is affected by additive Gaussian noise with variance $0.3^2$ while the initial times of the red, green, and blue car are $-5$, $5$, and $2$, respectively.}
\label{table:3}
\end{table}

\begin{remark}
In this case, the dimension of $x$ is $n=6$ as each of the three cars consists of two states. As mentioned in Remark \ref{rem:comp_compl}, we can reduce the computational complexity for calculating $\theta^c(x)$ under the assumption that the states of each car are subject to the same time shifts. In other words, the states of each car can be grouped and shifted synchronously so that the complexity is $O(r^3)$ instead of $O(r^6)$, which is a significant reduction. This way, the computation of $\theta^c(X^i)$ for a realization $X^i$ only took around $0.25$  s on a 1,4 GHz Intel Core i5.
\end{remark}

\subsection{Cooperative Servicing}

\begin{figure}[t]
	\centering
	\includegraphics[scale=0.475]{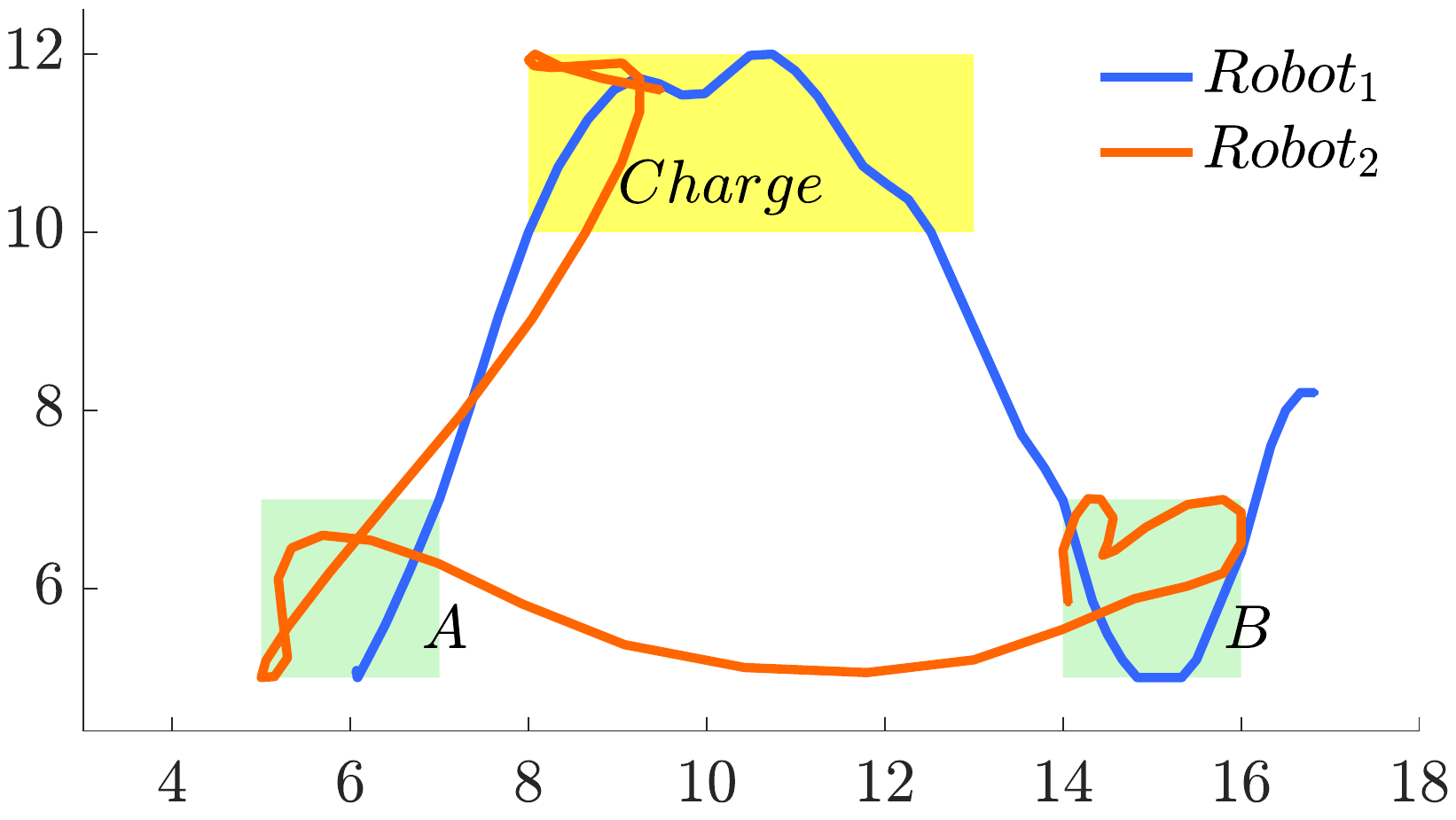}
	\caption{Nominal trajectories of the two robots under the control policies $u_i(t)$, i.e., system \eqref{eq:nominal_sims} without any time delays. The orange and blue stars denote initial conditions.}
	\label{ex:drone_setup}
\end{figure}

Consider two mobile robots employed in a  cooperative  servicing mission, see Figure \ref{ex:drone_setup} for an overview. The robots operate in a two-dimensional workspace and are described by the state $x_{\{i\}}$ consisting of its position $p_i$ and its velocity $v_i$, i.e., $x_{\{i\}}:=(p_i,v_i)\in\mathbb{R}^4$ for robot $i$. The state $x:=(x_{\{1\}},x_{\{2\}})$ is driven by discrete-time double integrator dynamics 
\begin{align}\label{eq:nominal_sims}
x_{\{i\}}(t+1):=Ax_{\{i\}}(t)+Bu_i(t)
\end{align}
where $u_i:\mathbb{N}\to \mathbb{R}^2$ is the control input and  where $A$ and $B$ are the discretized system and input matrices for two-dimensional double integrator dynamics under sampling time $t_s:=0.1$. Assume now that we are given the STL  specification
\begin{subequations}
\begin{align}
\phi&:=F_{[0,1]}\big((p_1\in A \vee p_2 \in A) \wedge F_{[1,5]}(p_1\in A \vee p_2 \in A)\big)\label{eq_sim_1}\\
& \wedge F_{[1,6]}(p_1\in B \wedge p_2 \in B)\label{eq_sim_2}\\
& \wedge F_{[0,2]}G_{[0,0.5]} (p_1\in \text{Charge}) \wedge F_{[0,2]}G_{[0,0.5]} (p_2\in \text{Charge}).\label{eq_sim_3}
\end{align}
\end{subequations}

The first line \eqref{eq_sim_1} encodes a sequencing task, i.e., eventually within $1$ time unit either robot $1$ or robot $2$ is in region $A$, followed by eventually within $[1,5]$ time units either robot $1$ or robot $2$ is again in region $A$. The second line \eqref{eq_sim_2} encodes that eventually within $[1,6]$ time units both robots meet in region $B$. The third line \eqref{eq_sim_3} encodes that  both robots eventually within $2$ time units visit the charging areas for at least $0.5$ time units. 

We first consider the deterministic system in \eqref{eq:nominal_sims} and synthesize a sequence of control inputs $u_i(t)$ for each robot $i$ following the approach presented in \cite{rodionova2021time}. For the deterministic system, we have that $\eta^\phi(X,0)=7$ time steps and  $\theta^\phi(X,0)=6$ time steps. Note here the particular interplay between  time units and the sampling time $t_s$, i.e., each time unit consists of $10$ time steps.

We now delay the initial time of each robot by $d_i$ similarly to the previous example, but here $d_i$ follows a Poisson distribution with probability density function $\frac{\lambda_i^{d_i}\exp(-\lambda)}{d_i\,!}$ and parameter $\lambda_i>0$.  Our goal is now to verify the STL temporal robustness risk under the sequence of control inputs $u_i$. The results for $\eta^\phi(X,0)$ are presented in Table \ref{table:4}  where $\lambda_i$ corresponds to the Poisson parameter for $d_i$.
\begin{table}[h!]
	\centering
\begin{tabular}{|p{2.5cm}|p{1.25cm}|p{1.25cm}|p{1.25cm}|p{1.25cm}|p{0.6cm}|}
	\hline
	\backslashbox{}{Risk} & $\overline{VaR}_{0.85}$ & $\overline{VaR}_{0.9}$ & $\overline{VaR}_{0.95}$ & $\overline{VaR}_{0.98}$  & $\#$\\
	\hline
	$\lambda_1=\lambda_2=1$ & -7 & -7 & -7 & -6 &  0\\
	$\lambda_1=\lambda_2=3$  & -5 & -5 & -4 & -2 &  1 \\
	$\lambda_1=\lambda_2=5$  & -3 & -2 & -1 & 1 &  139 \\
	$\lambda_1=\lambda_2=7$  & 0 & 0 & 1 & 4 &  1002 \\
	\hline
	$\lambda_1=1$, $\lambda_2=5$  & -7 & -7 & -7 & -6  &  9 \\
	$\lambda_1=5$, $\lambda_2=1$  & -3 & -2 & -1 & 1  &  137 \\
	\hline
\end{tabular}
\caption{Risk estimates of $-\eta^\phi(X,0)$ where the initial  time of each robot is Poisson distributed with parameters $\lambda_i$.}
\label{table:4}
\end{table}

Note here that similar general observation as in the T-intersection scenario can be made, i.e., lower risks are preferable so that naturally the cases with lower $\lambda_i$ are preferable. The results for $\theta^\phi(X,0)$ are presented in Table \ref{table:5}.
\begin{table}[h!]
	\centering
\begin{tabular}{|p{2.5cm}|p{1.25cm}|p{1.25cm}|p{1.25cm}|p{1.25cm}|p{0.6cm}|}
	\hline
	\backslashbox{}{Risk} & $\overline{VaR}_{0.85}$ & $\overline{VaR}_{0.9}$ & $\overline{VaR}_{0.95}$ & $\overline{VaR}_{0.98}$  & $\#$\\
	\hline
	$\lambda_1=\lambda_2=1$ & -5 & -5 & -4 & -4 &  0\\
	$\lambda_1=\lambda_2=3$  & -4 & -4 & -3 & -2 &  1 \\
	$\lambda_1=\lambda_2=5$  & -2 & -2 & -1 & 1 &  139 \\
	$\lambda_1=\lambda_2=7$  & 0 & 0 & 1 & 4 &  1002 \\
	\hline
	$\lambda_1=1$, $\lambda_2=5$  & -2 & -2 & -1 & 0  &  9 \\
	$\lambda_1=5$, $\lambda_2=1$  & -3 & -2 & -1 & 1  &  137 \\
	\hline
\end{tabular}
\caption{Risk estimates of $-\theta^\phi(X,0)$ where the initial  time of each robot is Poisson distributed with parameters $\lambda_i$.}
\label{table:5}
\end{table}

Tables \ref{table:4} and \ref{table:5} are according to Corollary \ref{cor:eta_theta} which can be shown to also hold for STL specifications $\phi$, i.e., $|\theta^\phi(X,0)|\le |\eta^\phi(X,0)|$. Notable  are the cases $\lambda_1=1$, $\lambda_2=5$  and  $\lambda_1=5$, $\lambda_2=1$. For the latter case, the bound $|\theta^\phi(X,0)|\le |\eta^\phi(X,0)|$ proves to be conservative and shows the sensititivity of $\theta^\phi(X,0)$  to asynchronous time shifts.

\section{Conclusion}
\label{sec:conclusion}
This paper studied temporal robustness of stochastic signals. We presented notions of temporal robustness for deterministic and stochastic signals. Our framework estimates the risk associated with a stochastic signal to not satisfy a specification robustly in time. We particularly permit general forms of risk measures to enable an axiomatic risk assessment. Case studies complemented the theoretical results highlighting the importance in applications and for risk-aware decision making and system verification.

\bibliographystyle{IEEEtran}
\bibliography{literature}

\appendix

\section{Proof of Lemma \ref{lem:1}}
    Assume that the synchronous temporal robustness is such that $|\eta^c(x)|=\zeta \geq 0$. By Corollary \ref{cor:1}, it follows that
\begin{equation}
\label{eq:beta_seq}
\forall \kappa\in[-\zeta,\zeta],\ \beta^c(x_\kappa)=\beta^c(x).
\end{equation}
Now, take any $\xi\in\mathbb{Z}$. By definition, we have
\begin{equation}
\label{eq:eta_xi}
\eta^c(x_\xi):=\beta^c(x_\xi) \sup\{\tau\in \mathbb{N}|\forall \kappa\in[-\tau,\tau], \beta^c(x_{\xi+\kappa})=\beta^c(x_\xi)  \}.
\end{equation}
First, assume that $\beta^c(x_\xi)=1$ so that $\eta^c(x_\xi)\geq 0$.
\begin{itemize}
	\item For $|\xi|> \zeta$,  it trivially follows that $\eta^c(x_\xi)\geq 0 > \zeta-|\xi|\ge \eta^c(x)-|\xi|$.
	\item For $|\xi|\le \zeta$,  it holds that $\beta^c(x)=\beta^c(x_\xi)=1$ due to \eqref{eq:beta_seq}.
	From \eqref{eq:beta_seq} it follows that $\forall \kappa\in[-\zeta+|\xi|,\zeta-|\xi|]$, $\beta^c(x_{\xi+\kappa})=\beta^c(x)=\beta^c(x_\xi)$ since $\xi+\kappa\in [\xi-\zeta+|\xi|,\xi+\zeta-|\xi|]\subseteq [-\zeta,\zeta]$. Together with the definition of $\eta^c(x_\xi)$ in \eqref{eq:eta_xi} this leads to $\eta^c(x_\xi) \geq \zeta-|\xi|\ge \eta^c(x)-|\xi|$.
\end{itemize}
Next, assume that $\beta^c(x_\xi)=-1$ so that $\eta^c(x_\xi)\leq 0$.
\begin{itemize}
	\item For $\beta^c(x)=1$, we have $\eta^c(x)=\zeta \ge 0$. This implies  $\forall \kappa\in[-\zeta,\zeta],\ \beta^c(x_\kappa)=1$ by \eqref{eq:beta_seq} so that, as we assume $\beta^c(x_\xi)=-1$, it has to hold that $|\xi|>\zeta$. Note that there exists a $\kappa$ with $|\kappa|=||\xi|-\zeta|$  so that $|\xi+\kappa|= \zeta$ and for which hence $\beta^c(x_{\xi+\kappa})=1$ so that $\eta^c(x_\xi)\ge -(||\xi|-\zeta|-1)\ge-(|\xi|-\zeta)=  \eta^c(x)-|\xi|$ by \eqref{eq:eta_xi}.
	\item For $\beta^c(x)=-1$, note that there exists $\kappa:=-\xi+b\pm 1$ where $|b|=  \zeta$ is such that  $\beta^c(x_{\xi+\kappa})=1$ so that $\eta^c(x_\xi)\ge -(|\xi|+\zeta)= \eta^c(x)-|\xi|$ by \eqref{eq:eta_xi}.\footnote{The intuition of $\kappa$ here is that one can shift $x_{\xi}$ to $x$ by $-\xi$ shifts first. There then has to exist a shift $b\pm1$ of $x$ so that $\beta^c(x_{\xi+\kappa})=1$.} 
\end{itemize}
It hence follows that $\eta^c(x_\xi) \ge \eta^c(x)-|\xi|$.

\section{Proof of Lemma \ref{lem:2}}
The proof follows similarly to the proof of Lemma \ref{lem:1}.     Assume that the asynchronous temporal robustness is such that $|\theta^c(x)|=\zeta \geq 0$. By Corollary \ref{cor:2}, it  follows that
\begin{equation}
\label{eq:beta_seq_}
\forall \kappa_1,\hdots,\kappa_n\in[-\zeta,\zeta],\ \beta^c(x_{\bar{\kappa}})=\beta^c(x).
\end{equation}
Now, take any $\xi_1,\hdots,\xi_n\in\mathbb{Z}$. By definition, we have
\begin{equation}
\label{eq:eta_xi_}
\theta^c(x_{\bar{\xi}}):=\beta^c(x_{\bar{\xi}}) \sup\{\tau\in \mathbb{N}|\forall \kappa_1,\hdots,\kappa_n\in[-\tau,\tau], \beta^c(x_{\bar{\xi}+\bar{\kappa}})=\beta^c(x_{\bar{\xi}})  \}
\end{equation}
where $\bar{\xi}+\bar{\kappa}:=(\xi_1+\kappa_1,\hdots,\xi_n+\kappa_n)$. For brevity, let us also denote $\xi^*:=\max(|\xi_1|,\hdots,|\xi_n|)$.

First, assume that $\beta^c(x_{\bar{\xi}})=1$ so that $\theta^c(x_{\bar{\xi}})\geq 0$.
\begin{itemize}
	\item For $\xi^*> \zeta$,  it trivially follows that $\theta^c(x_{\bar{\xi}})\geq 0 > \zeta-\xi^*\ge \theta^c(x)-\xi^*$.
	\item For $\xi^*\le \zeta$,  it holds that $\beta^c(x)=\beta^c(x_{\bar{\xi}})=1$ due to \eqref{eq:beta_seq_}.
	From \eqref{eq:beta_seq_} it follows that $\forall \kappa_i\in[-\zeta+\xi^*,\zeta-\xi^*]$, $\beta^c(x_{\bar{\xi}+\bar{\kappa}})=\beta^c(x)=\beta^c(x_{\bar{\xi}})$ since $\xi_i+\kappa_i\in [\xi_i-\zeta+\xi^*,\xi_i+\zeta-\xi^*]\subseteq [-\zeta,\zeta]$. Together with the definition of $\theta^c(x_{\bar{\xi}})$ in \eqref{eq:eta_xi_} this leads to $\theta^c(x_{\bar{\xi}}) \geq \zeta-\xi^*\ge \theta^c(x)-\xi^*$.
\end{itemize}
Next, assume that $\beta^c(x_{\bar{\xi}})=-1$ so that $\theta^c(x_{\bar{\xi}})\leq 0$.
\begin{itemize}
	\item For $\beta^c(x)=1$, we have $\theta^c(x)=\zeta \ge 0$. This implies that  $\forall \kappa_1,\hdots,\kappa_n\in[-\zeta,\zeta],\ \beta^c(x_{\bar{\kappa}})=1$ by \eqref{eq:beta_seq_} so that, as we assume $\beta^c(x_{\bar{\xi}})=-1$, it has to hold that $\xi^*>\zeta$. Note that there exists $\kappa_1,\hdots,\kappa_n$ with $|\kappa_i|\le |\xi^*-\zeta|$  so that $|\xi_i+\kappa_i|\le \zeta$ and for which hence $\beta^c(x_{\bar{\xi}+\bar{\kappa}})=1$ so that $\theta^c(x_{\bar{\xi}})\ge -(|\xi^*-\zeta|-1)\ge-(\xi^*-\zeta)=  \theta^c(x)-\xi^*$ by \eqref{eq:eta_xi_}.
	\item For $\beta^c(x)=-1$, note that there exists $\kappa_i:=a_i+b_i\pm 1$ where $|a_i|\le \xi^*$ and $|b_i|\le   \zeta$ such that  $\beta^c(x_{\bar{\xi}+\bar{\kappa}})=1$ so that $\theta^c(x_{\bar{\xi}})\ge -(\xi^*+\zeta)= \theta^c(x)-\xi^*$ by \eqref{eq:eta_xi_}.
\end{itemize}
It hence follows that $\theta^c(x_\xi) \ge \theta^c(x)-\xi^*$.

\section{Proof of Corollary \ref{cor:eta_theta}}
	Assume that the asynchronous temporal robustness is such that $|\theta^c(x)|=\zeta \geq 0$. By Corollary \ref{cor:2}, we again have  that \eqref{eq:beta_seq_} holds. Therefore,  $\beta^c(x_{\bar{\kappa}})=\beta^c(x)$ if $\kappa_1=\hdots=\kappa_n$ and $\kappa_1\in[-\zeta,\zeta]$. Consequently, it holds  $\beta^c(x_{\kappa})=\beta^c(x)$ for all $ \kappa\in[-\zeta,\zeta]$. By the definition of $\eta^c(x)$, it follows that 
$|\eta^c(x)|\geq \zeta=|\theta^c(x)|$.

\section{Proof of Theorem \ref{thm:1}}
 \emph{Step 1 - Measurability of $\beta^c(X(\cdot,\omega))$:} Recall that
\begin{align*}
\beta^c(X(\cdot,\omega))=\begin{cases}
1 & \text{if } \inf_{t\in\mathbb{Z}} c(X(t,\omega),t)\ge 0 \\
-1 & \text{otherwise}
\end{cases}
\end{align*}
where $c(X(t,\cdot),t)$ is a random variable as noted before, i.e., the function $c(X(t,\omega),t)$ is measurable in $\omega$ for each fixed $t\in\mathbb{Z}$.    It holds that $\inf_{t\in\mathbb{Z}} c(X(t,\omega),t)$ is measurable in $\omega$ as the infimum over a countable number of measurable functions is again measurable, see e.g., \cite[Theorem 4.27]{guide2006infinite}. Note next that the characteristic function of a measurable set is measurable again (see e.g., \cite[Chapter 1.2]{durrett2019probability}) so that $\beta^c(X(\cdot,\omega))$ is  measurable in $\omega$.

\emph{Step 2 - Measurability of $\eta^c(X(\cdot,\omega))$:}  First, let $X_\kappa(t,\omega):=X(t+\kappa,\omega)$ in direct analogy to $x_\kappa$ defined previously. It holds that $X_\kappa(t,\omega)$ is measurable in $\omega$ since $X(t+\kappa,\omega)$ is measurable in $\omega$.    Recall that
\begin{align*}
\eta^c(X(\cdot,\omega))&:=\beta^c(X(\cdot,\omega))\sup\{\tau\in \mathbb{N}|\forall \kappa\in[-\tau,\tau], \beta^c(X_\kappa(\cdot,\omega))=\beta^c(X(\cdot,\omega))  \}
\end{align*}
Next, let us rewrite the supremum within the previous expression  as $\sup_{\tau\in\mathbb{N}} g_\tau(X(\cdot,\omega))$ where each $g_\tau(X(\cdot,\omega))$ is such that 
\begin{align*}
g_\tau(X(\cdot,\omega))=\begin{cases}
\tau &\text{if } \forall \kappa\in[-\tau,\tau], \beta^c(X_\kappa(\cdot,\omega))=\beta^c(X(\cdot,\omega))\\
0 &\text{otherwise}.
\end{cases}
\end{align*}
In particular, the function $g_\tau$ is equivalent to
\begin{align*}
g_\tau(X(\cdot,\omega)):=\begin{cases}
\tau & \text{if } \sum_{\kappa=-\tau}^\tau |\beta^c(X_\kappa(\cdot,\omega))-\beta^c(X(\cdot,\omega))|=0\\
0 & \text{otherwise}.
\end{cases}
\end{align*}
Since $\sum_{\kappa=-\tau}^\tau |\beta^c(X_\kappa(\cdot,\omega))-\beta^c(X(\cdot,\omega))|$ is measurable in $\omega$ and since the characteristic function of a measurable set is again measurable, it holds that $g_\tau(X(\cdot,\omega))$ is measurable in $\omega$. It follows that  $\sup_{\tau\in\mathbb{N}} g_\tau(X(\cdot,\omega))$ is measurable in $\omega$ as the supremum over a countable number of measurable functions is  measurable, see e.g., \cite[Theorem 4.27]{guide2006infinite}. It consequently holds that $\eta^c(X(\cdot,\omega))$ is measurable in $\omega$. 

\emph{Step 3 - Measurability of $\theta^c(X(\cdot,\omega))$:}  Measurability of $\theta^c(X(\cdot,\omega))$ can be shown similarly to measurability of $\eta^c(X(\cdot,\omega))$ by changing the definition of  $g_\tau$ to include a nested sum over $\kappa_1,\hdots,\kappa_n$.

\section{Proof of Theorem \ref{thm:2}}
		Every realization $X(\cdot,\omega)$ for $\omega\in\Omega$ leads to a synchronous temporal robustness  $\eta^c(X(\cdot,\omega))$. Additionally, every realization $\xi(\omega)$   for $\omega\in\Omega$ leads to a synchronous temporal robustness  $\eta^c(X_{\xi}(\cdot,\omega))$.\footnote{Recall that $X_{\xi}(t,\omega)=X(t+\xi(\omega),\omega)$. Note that it can be shown that $X_{\xi}(\cdot,\omega)$ is measurable in $\omega\in\Omega$.}  According to Lemma \ref{lem:1} and since $|\xi(\omega)|\le d$, it holds that 
		\begin{align*}
		\eta^c(X_{\xi}(\cdot,\omega))\ge \eta^c(X(\cdot,\omega))-|\xi(\omega)|\ge\eta^c(X(\cdot,\omega))-d
		\end{align*} 
		so that  $-\eta^c(X_\xi(\cdot,\omega))\le -\eta^c(X(\cdot,\omega))+d$. As $R$ is monotone and  translationally invariant, it follows that 
		\begin{align*}
		R(-\eta^c(X_\xi))\le R(-\eta^c(X)+d)= R(-\eta^c(X))+d.
		\end{align*}
		
\section{Proof of Theorem \ref{thm:3}}
		The proof follows similarly to the proof of Theorem \ref{thm:2}. According to Lemma \ref{lem:2} and as we have that $\max(|\xi_1(\omega)|,\hdots,|\xi_n(\omega)|)\le d$, it holds that
		\begin{align*}
		\theta^c(X_{\bar{\xi}}(\cdot,\omega))\ge \theta^c(X(\cdot,\omega))-d
		\end{align*}
		so that  $-\theta^c(X_{\bar{\xi}}(\cdot,\omega))\le -\theta^c(X(\cdot,\omega))+d$. As $R$ is monotone and  translationally invariant, it follows that 
		\begin{align*}
	R(-\theta^c(X_{\bar{\xi}}))\le R(-\theta^c(X)+d)= R(-\eta^c(X))+d.
	\end{align*}

\addtolength{\textheight}{-12cm}   

\end{document}